\documentclass[10pt,twocolumn,letterpaper]{article}

\usepackage{cvpr}              %

\usepackage[dvipsnames]{xcolor}
\usepackage{amsmath}
\usepackage{amssymb}
\usepackage{amsfonts}
\usepackage{amsthm}
\usepackage{bbm}
\usepackage[linesnumbered,ruled,vlined]{algorithm2e}

\SetKwInput{KwInput}{Input}                %
\SetKwInput{KwOutput}{Output}  
\newcommand{\Dr}{{D_{\text{retr}}}}
\newcommand{\Ds}{D_{\text{store}}}
\newcommand{\Dp}{D_{\text{private}}}
\newcommand{\safe}{\operatorname{safe}}

\newtheorem{theorem}{Theorem}
\newtheorem{lemma}{Lemma}

\newtheorem{definition}{Definition}
\newtheorem{proposition}{Proposition}

\renewcommand{\paragraph}[1]{\vspace{0.5em}\noindent\textbf{#1}}

\definecolor{cvprblue}{rgb}{0.21,0.49,0.74}
\usepackage[pagebackref,breaklinks,colorlinks,citecolor=cvprblue]{hyperref}
\usepackage[accsupp]{axessibility} %

\def\paperID{13987} %
\def\confName{CVPR}
\def\confYear{2024}

\title{CPR: Retrieval Augmented Generation for Copyright Protection}

\author{%
  Aditya Golatkar \quad Alessandro Achille \quad Luca Zancato \\
  Yu-Xiang Wang \quad Ashwin Swaminathan \quad Stefano Soatto \\
  AWS AI Labs\\
  \texttt{\{agolatka,aachille,zancato,yuxiangw, swashwin,soattos\}@amazon.com} \\
}

\begin{document}
\maketitle
\begin{abstract}
Retrieval Augmented Generation (RAG) is emerging as a flexible and robust technique to adapt models to private users data without training, to handle credit attribution, and to allow efficient machine unlearning at scale. However, RAG techniques for image generation may lead to parts of the retrieved samples being copied in the model's output. To reduce risks of leaking private information contained in the retrieved set, we introduce Copy-Protected generation with Retrieval (CPR), a new method for RAG with strong copyright protection guarantees in a mixed-private setting for diffusion models. 
CPR allows to condition the output of diffusion models on a set of retrieved images, while also guaranteeing that unique identifiable information about those example is not exposed in the generated outputs. In particular, it does so by sampling from a mixture of public (safe) distribution and private (user) distribution by merging their diffusion scores at inference.
We prove that CPR satisfies Near Access Freeness (NAF) which bounds the amount of information an attacker may be able to extract from the generated images. We provide two algorithms for copyright protection, CPR-KL and CPR-Choose. Unlike previously proposed rejection-sampling-based NAF methods, our methods enable efficient copyright-protected sampling with a single run of backward diffusion.
We show that our method can be applied to any pre-trained conditional diffusion model, such as Stable Diffusion or unCLIP. In particular, we empirically show that applying CPR on top of unCLIP improves quality and text-to-image alignment of the generated results (81.4 to 83.17 on TIFA benchmark), while enabling credit attribution, copy-right protection, and deterministic, constant time, unlearning.

\end{abstract}

\begin{figure*}[t]
\centering
\includegraphics[width=0.9\textwidth]{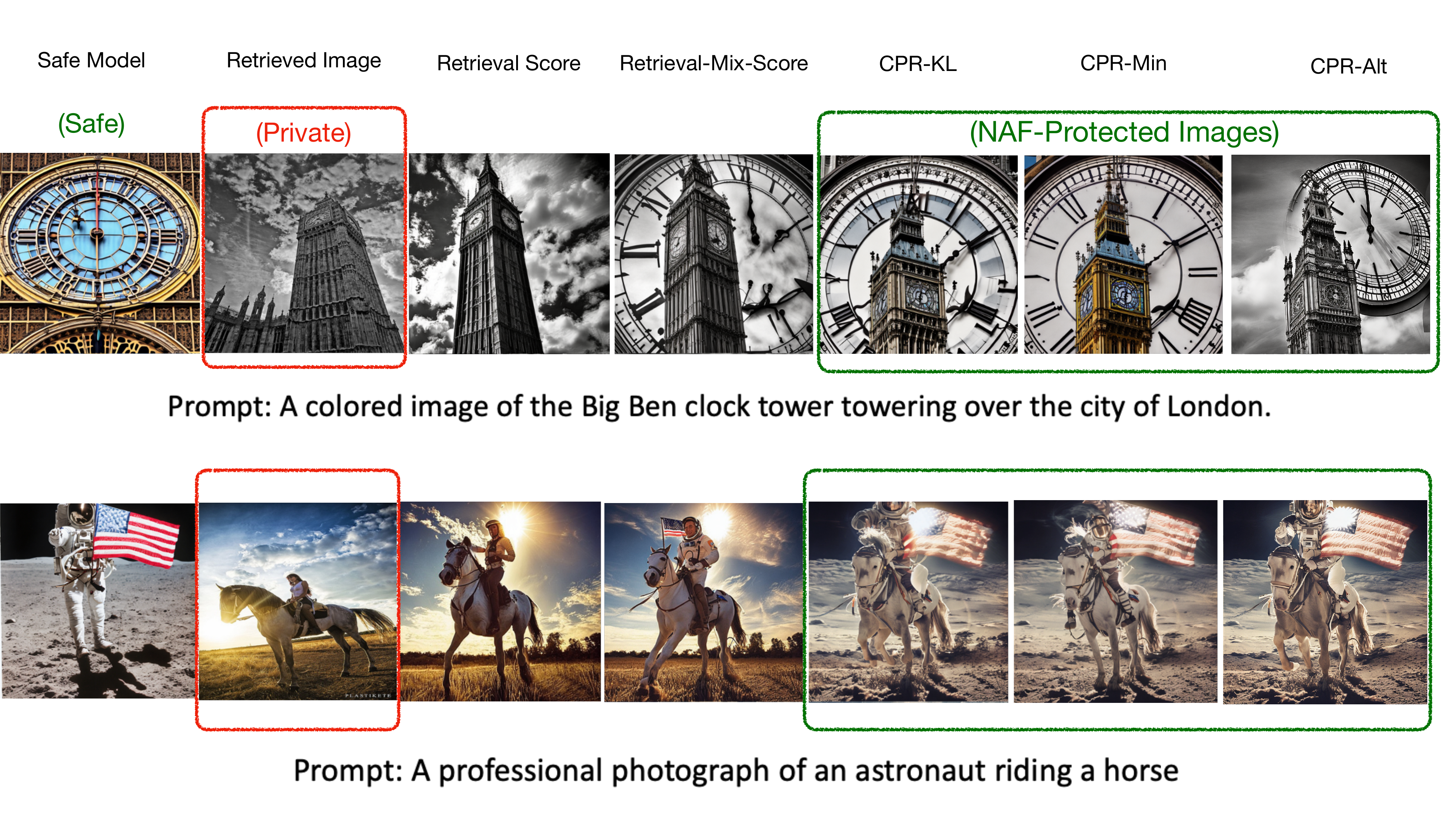}
\vspace{-13pt}
\caption{\textbf{RAG vs CPR image generation.}
Images generated using the given prompt for a fixed random seed using different methods. \textit{Safe Model}: Pre-trained model with no access to the retrievable data store, \textit{Retrieval-Score}: Image generated using \cref{eq:retrieval-score}, \textit{Retrieval-Mix-Score}: Image generated using \cref{eq:retrieval-score-complete}, \textit{CPR-KL, CPR-Min, CPR-Alt}: Images generated using our algorithms in \cref{algo:cpr-kl} \cref{cpr-min} and \cref{algo:cpr-int} \cref{cpr-alt}. Images generated without CPR bear more resemblance to the retrieved image, compared to the CPR generated images, which are different from the retrieved image, while preserving the underlying concept in the prompt (for example the astronaut seems to be on Moon, Big Ben is more textured with different design).}
\label{fig:splash-figure}
\end{figure*}

\vspace{-12pt}
\section{Introduction}
\label{sec:intro}
Foundation model users may need to adapt large-scale Diffusion Models to their use cases, like personalization, editing, content creation etc. However, fine-tuning the model on the user data is often not an option. This is in part due to the steep cost of fine-tuning models, but also because user data is a mutable entity: new data is constantly added, and low-quality data may often be filtered out. Moreover, data owners may, at any point, change their mind and demand that their data be removed.

Retrieval Augmented Generation (RAG) has emerged as a promising method to handle these situations. Rather than using the user data to fine-tune the model, supporting samples are retrieved from it at inference time to guide the generation of new samples. Data may be easily added or removed from the retrieval data store without changes to the model, and users may access different subsets of the data based on their access-right. 
However, RAG methods are double edged, direct access to retrieved reference images often significantly improves the quality of generated samples but as we depict in \cref{fig:splash-figure}, RAG models are prone to copy information from the retrieved examples into the model output, potentially resulting in significant leak of private information. 
We formalize this observation in \cref{sec:cpr} and we show that applying RAG on top of a public model, while retrieving private user data at inference time, cannot satisfy even the weaker notion of privacy. 

To remedy this, we introduce Copy-Protected Generation with Retrieval (CPR). CPR retrieves multiple private examples from the private user data pool. Information from all these samples is combined to generate a ``private'' diffusion flow which uses common information of those samples while discarding any unique and identifiable information. The resulting private flow is then optimally combined with the ``public'' flow generated by the base model to generate new outputs which still benefit from the retrieved samples, but minimize the risk of information leak.

In particular, we show that our method satisfies the recently proposed notion of copyright protection using Near Access Freeness (NAF) \cite{vyas2023provable}, a relaxation of differential privacy aimed at protecting specific attribute of the training data. 
Differently from previously proposed methods like \cite{vyas2023provable} that realize NAF with a computationally expensive rejection sampling method, CPR does so by construction during the generation. Hence making our method significantly faster than the previous baselines and while also keeping inference cost deterministic.

Theoretically, we formally prove in \cref{lemma:cpr-kl,lemma:choose} that CPR offers strong protection guarantees by ensuring that the generated samples contain at most $k$-bits of unique information about retrieved samples, where $k$ can be tuned by the user as required by the application.
Empirically, we show that CPR can use private data to improve quality of the generated images ($81.4 \to 83.17$ TIFA score) while maintaining privacy guarantees on the retrievable data.

The rest of the paper is organized as follows. In \cref{sec:related-works} we provide a study of the relevant related works in RAG and privacy. In \cref{sec:preliminaries} we define the necessary notations, and in \cref{sec:RAG} show how to perform RAG with pre-trained text-to-image diffusion models by formulating inference over mixture of public-private distribution. In \cref{sec:cpr} we provide our CPR algorithm, along with its theoretical guarantees, provide empirical evaluation of our method in \cref{sec:exps}, followed by some discussion in \cref{sec:disc}.

\vspace{-5pt}
\section{Related Works}
\label{sec:related-works}
\vspace{-5pt}
\paragraph{Retrieval Augmented Generation}
Retrival Augmented Generation (RAG) methods have been successfully applied to large language models (LLMs) \cite{khandelwal2019generalization,shi2023replug,ram2023context,guu2020retrieval,min2023silo}. RAG has been shown to outperform even LLMs trained jointly on the training set and the retrievable data pool. RAG  have also been explored for image synthesis \cite{blattmann2022retrieval,rombach2022text,zhang2023remodiffuse,chen2022re,yasunaga2023retrieval,sheynin2022knn,chen2023label}. However, rather than reusing existing models, current methods require training of retrieval-specific architectures which --- unlike the standard text-to-image diffusion models \cite{rombach2022high,ramesh2021zero,ramesh2022hierarchical,yu2022scaling} --- can be prompted with several retrieved images along with conditional information, such as text, as inputs. Instead, we explore RAG using more generic pre-trained text-to-image diffusion models. \cite{blattmann2022retrieval,rombach2022text,zhang2023remodiffuse} train a diffusion-based image retrieval model that can be prompted with latent image embeddings, while \cite{chen2022re,yasunaga2023retrieval} use autoregressive generative models inspired from LLMs. 

\paragraph{Image Manipulation}
Several recent works \cite{gal2022image,kumari2023multi,ruiz2023dreambooth,lu2023tf,brooks2023instructpix2pix,hertz2022prompt,ma2023subject,wang2023compositional} have provided methods for image manipulation, editing by either fine-tuning or changing the cross-attention values at inference. With an appropriate retrieval function, and database such methods can be used to perform retrieval augmented generation by merging diffusion scores \cite{golatkar2023training,du2023reduce,liu2022compositional}. However, the manipulation methods significantly lower inference speed. Instead, we opt to use the unCLIP model \cite{ramesh2022hierarchical} to generate a backward flow using the retrieved images, and merge it with the flow generated by a base text-to-image diffusion model at inference \cite{golatkar2023training,liu2022compositional,du2023reduce,rombach2022high}. 

\paragraph{Privacy}
Recent works  \cite{carlini2021extracting,carlini2023extracting,somepalli2023diffusion,somepalli2023understanding} have shown that such models are able to memorize their training data. This raises several privacy challenges, including:

\textit{Unlearning:} Machine unlearning \cite{golatkar2020eternal,bourtoule2021machine} enables users to delete their data from the weights of trained models  \cite{fabbrini2020right, nakashima2020legal}.
\cite{dukler2023safe,golatkar2023training,bourtoule2021machine,gandikota2023erasing,kumari2023ablating,liu2023tangent} provide training methods which makes unlearning efficient, for example by breaking the dataset into multiple shards and training separate models on each, followed by ensembling at inference. Despite their improved privacy utility trade-offs compared to a single model, such approaches still require frequent re-training/fine-tuning. On the other hand, we propose splitting the training dataset into a core safe dataset \cite{golatkar2021mixed} used to train a core model, and a user owned private data store used to retrieve samples. This allow instantaneous forgetting of any private sample without having to retrain the model. This strategy also improve performance (alignment), and enables easy continual learning by simply adding new data to the data store. Moreover, privacy of data which are never retrieved is completely preserved, unlike unlearning or differential privacy \cite{dwork2006differential,golatkar2022mixed} methods which mix information about the entire dataset in the weights.

\textit{Copyright Protection:} Memorization in foundation models also increases the risk of copying, style mimicry and copyright \cite{carlini2021extracting,carlini2023extracting,shan2023glaze} at inference.  \cite{vyas2023provable} proposed a definition for copying in generative models using near access freeness (NAF), and provided the CP-$\Delta$ algorithm for copy-protected generation. CP-$\Delta$ uses two generative models, trained on two disjoint splits of the data, and then at inference samples from the product and the minimum of the two distributions. However, using it directly out-of-the-box for diffusion models is challenging. Instead, they propose another algorithm, CP-$k$, based on rejection sampling. Diffusion models however tend to have slow inference speed, and adding rejection sampling further aggravates the speed problem. To address this, we introduce CPR, which provides a method to sample using CP-$\Delta$ (satisfies NAF) in a single run, without the need for rejection sampling.

\section{Preliminaries}
\label{sec:preliminaries}

Let $p_0(x_0)$ be a data distribution over images, which we seek to model using a diffusion model \cite{song2020score,song2020denoising,ho2020denoising}. Score based diffusion models models \cite{song2020score} define a (variance preserving) forward flow through a SDE, which transforms the distribution $p_0(x_0)$ at time $t=0$ in a reference distribution $p_1(x_1) = N(0, I)$ at time $t=1$:
\begin{equation}
    dx_t = -\dfrac{1}{2} \beta_t  x_tdt + \sqrt{\beta_t}d\omega_t
\label{eq:forward-process}
\end{equation}
where $x_t$ is the diffused input at time $t$, $d\omega_t$ is a standard Wiener process, and $\beta_t$ are time varying coefficients (in practice implemented through linear or cosine scheduling), which determine the transition kernel and amount of noise added over time. 
The intermediate result $p_t(x_t)$ of the diffusion process at time $t$  equivalently expressed as the result of applying a Gaussian kernel $p_t(x_t|x_0) = \mathcal{N}(x_t;\gamma_tx_o, \sigma^2_tI)$ to $p_0(x_0)$, resulting in $p_t(x_t) = \int_{x_0} p_t(x_t|x_0) p_0(x_0) dx_0$, where $\gamma_t = \exp(-\frac{1}{2} \cdot \int_0^{t}\beta_tdt)$ and $\sigma^2_t = 1 - \gamma^2_t$.

The forward process in \cref{eq:forward-process} can be inverted through a corresponding backward process \cite{lindquist1979stochastic, song2020score}. In particular, this process can be used to generate samples of $p_0(x_0)$ starting from a sample of $p_1(x_1)=N(0, I)$:
\begin{equation}
dx_t = \big(- \dfrac{1}{2}\beta_t  x_t - \nabla_{x_t} \log p_t(x_t)\big)dt + \sqrt{\beta_t}d\omega_t
\label{eq:reverse-process-sde}
\end{equation}
where $\nabla_{x_t} \log p_t(x_t)$ is the score function of data distribution at $t$. Efficiently computing the score function is difficult. Instead, it can be approximated  $\nabla_{x_t} \log p_t(x_t) \approx s_{\theta}(x_t, t)$ using a deep network $s_{\theta}(x_t, t)$, \ie, a diffusion model. %
In practice, diffusion models are often trained to take additional inputs $s_{\theta}(x_t, t, c)$  in order to model a conditional distribution $p_0(x_0|c)$, where the conditioning $c$ provides additional information about the sample to generate, such as textual prompts \cite{dhariwal2021diffusion,ho2022classifier}.
Given samples of the joint distribution $p_0(x_0,c)$, a diffusion model $s_{\theta}(x_t, t, c)$ can be trained by minimizing the score-matching objective:
\begin{equation}
\mathbb{E}_{(x_0,c) \sim p_0(x_0,c)}\mathbb{E}_t\big[\|s_{\theta}(x_t, t, c) - \nabla_{x_t}\log p_t(x_t|x_0)\|\big]
\label{eq:conditional-optimization}
\end{equation}
Directly generating samples using the backward flow modeled by $s_{\theta}(x_t, t, c)$ can result into poor alignment \cite{du2023reduce,ho2020denoising,song2020denoising}. This can be improved through classifier-free guidance \cite{ho2022classifier}, which uses the modified score: 
\[s_{\theta}(x_t, t, \phi) + \lambda (s_{\theta}(x_t, t, c) - s_{\theta}(x_t, t, \emptyset)),\]
where the hyper-parameter $\lambda$ controls the \textit{guidance scale}, and $\emptyset$ denotes that no conditioning is fed to the model.

\begin{table*}[t]
\centering
\includegraphics[width=0.95\textwidth]{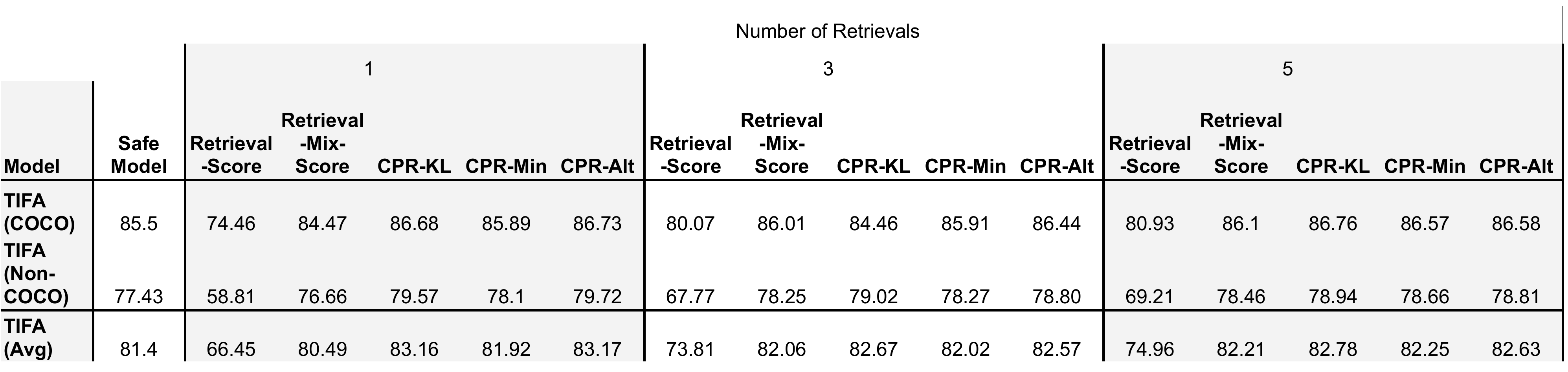}
\caption{\textbf{Improved text-to-tmage alignment with retrieval}: We compute the TIFA score \cite{hu2023tifa} which measures the text-to-image alignment on a set of prompts (Higher is better). We use a subset of MSCOCO \cite{lin2014microsoft} (2k images with high aesthetic score) as the private data store. We show that simply using the retrieval-score (in \cref{eq:retrieval-score}) is not enough to improve alignment, instead using the retrieval-mixture-score (in \cref{eq:retrieval-score-complete}) is important to generate aligned and well composed images. CPR-KL, CPR-Min, CPR-Alt, meaningfully improve the text-to-image alignment across different retrieval settings compared to the base model while protecting the private data store.}
\label{fig:tifa-table}
\end{table*}

\section{Mixed-Privacy RAG}
\label{sec:RAG}
In this section, we introduce a method for privacy-enabled RAG that is based on the notion of mixed-privacy \cite{golatkar2021mixed,golatkar2022mixed}. 

Let $D=\{x^i, c^i\}_{i=1}^{N} \sim p(x, c)$ be a safe training dataset --- meaning that samples are considered public in the differential-privacy sense (see \cite{golatkar2021mixed,golatkar2022mixed}). 
We assume $D$ is used to train a core public diffusion model $s_{\theta}(x_t, t, c)$, that accepts $c$ as conditioning information. We shall also assume that $c$ is the output of a CLIP encoder $c = \operatorname{CLIP}(\texttt{<prompt>})$ fed with either a text prompt or an image prompt.
Furthermore, let $\Dp=\{x^i, c^i\}_{i=1}^{M}$ be a private dataset which may require frequent unlearning, or may require privacy or copyright protection. 
We shall consider $\Dp$ as our data store for retrieval.

\paragraph{Retrieval}
At inference time, given a user prompt $c_{\text{test}}$ we retrieve a set of $m$ relevant examples $\Dr = \{(x_i, \phi(c_i, c_\text{test})\}_{i=1}^m \subset \Dp$ to aid the generation process. 
For simplicity, we simply retrieve the closest $m$ samples based on $L_2$-CLIP similarity score:
\[
\text{score} = \|c_\text{test} - c_i\| + \|c_\text{test} - \operatorname{CLIP}(x_i)\|.
\]
Note however that in $\Dr$ we are modifying the prompt of the retrieved samples through a function $\phi(c_i, c_\text{test}) = c_i + c_\text{test}$ in order to align them better with the user prompt.

\paragraph{Mixture-of-Distribution}
Retrieved samples are used to improve the generation of new samples.
Formally, the goal of CPR is to modify the sampling backward process in order to generate samples from a weighted mixture of the distribution of $D$ and $\Dr$ \cite{golatkar2023training,liu2022compositional,du2023reduce}:
\begin{equation}
p(x |c) = w_0p_{D}(x|c) + w_1p_{\Dr}(x|c) 
\label{eq:mixture-of-distribution}
\end{equation}
where the weights $w_0 = \lambda$ and $w_1=1 - \lambda$ allow the user to control the  contribution of the retrieved samples at inference time through an hyperparameter $0 < \lambda < 1$.

\paragraph{Mixture-of-Score} To sample from this mixture distribution, we need to compute its score function $\nabla \log p_t(x_t)$ at time $t$ (see eq.~\ref{eq:reverse-process-sde}). From \cref{sec:preliminaries} we have:
\begin{align}
p_t(x_t|c) = \int p_t(x_t|x_0)\big[w_0p_D(x_0|c) + w_1p_{D_\text{retr}}(x_0|c)\big]dx_0
\label{eq:mixture-at-t}
\end{align}
where $p_t(x_t|x_0) = \mathcal{N}(x_t;\gamma_tx_0, \sigma^2_tI)$ is a Gaussian kernel.
The following proposition expresses the score of the mixture as a function of the score of the individual components:
\begin{proposition}
\label{prop:mixture-of-score}
Let $p_t(x_t|c)$ be as in \cref{eq:mixture-at-t}, then  $\nabla_{x_t}\log p_t(x_t|c)$ is given by:
\begin{align}
\nabla_{x_t}\log p_t &= \hat{w}^t_0 \nabla_{x_t} \log p^t_D(x_t|c) + \hat{w}^t_1 \nabla_{x_t} \log p^t_{D_\text{retr}}(x_t|c) \nonumber
\end{align}
where we have defined:
\begin{align}
    \hat{w}^t_0&=w_0\dfrac{p^t_D(x_t|c)}{p_t(x_t|c)},\ \ 
    \hat{w}^t_1=w_1\dfrac{p^t_\Dr(x_t|c)}{p_t(x_t|c)}. \nonumber
\end{align}
and $p^t_D(x_t|c)$ denotes the forward flow of the distribution $p_D(x_t|c)$ at time $t$ (and similarly for $p^t_\Dr(x_t|c)$) and $p_t(x_t|c) = p^t_D(x_t|c) + p^t_\Dr(x_t|c)$.
\end{proposition}

While $\hat{w}^t_0$ and $\hat{w}^t_1$ could be computed exactly, we find that treating them as fixed hyper-parameters simplifies the implementation and performs well. The scores $\nabla_{x_t} \log p_D(x_t|c)$ can be approximated empirically by a diffusion model $s_{\theta_0}(x, t, c)$ trained  on $D$. However, we do not have a model trained on the retrieved data $\Dr$ to estimate $\nabla_{x_t} \log p_\Dr(x_t|c)$. To solve the issue, recall that such a diffusion model $s_{\theta_1}$ that minimizes the loss:
\begin{align}
&s_{\theta_1} = \arg\min_{s_{\theta}}\mathbb{E}_{(x_0,c) \sim p_{\Dr}}\mathbb{E}_{x_t} \big[ \| s_{\theta}(x_t, t, c) \nonumber \\
&\hspace{3em}- \nabla_{x_t} \log \big(\int p_t(x_t|x_0)p_{\Dr}(x_0, c) \big)\big\|\big]
\label{eq:loss-1}
\end{align}
Since $|\Dr| \ll |D|$, we expect the minimizer $\theta_1$ to be a small small perturbation $\theta_1 = \theta_0 + \Delta \theta_1$. However, fine-tuning $s_{\theta_0}(x, t, c)$ to find such $\Delta \theta_1$ for every inference request is computationally prohibitive.

Instead of fine-tuning, we approximate the expected behavior of $s_{\theta_1}$ through prompting. Textual inversion and prompt tuning  have been shown to perform comparably to fine-tuning on small datasets while using orders of magnitute less parameters \cite{gal2022image,kumari2023multi,wen2023hard,sohn2023visual}. However, despite the reduction, it is still cumbersome to fine-tune at inference. Instead we propose manually modifying the user prompt $c_\text{test}$ using the CLIP embeddings of the retrieved samples, and define the \textit{retrieval-score function}:
\begin{align}
\hat{s}_{\theta_0}(x_t, t, c_{\text{test}}) \triangleq s_{\theta_0}\Big(x_t, t, \frac{1}{m} \sum_{x_i \in \Dr}\text{CLIP}(x_i)\Big)
\label{eq:retrieval-score}
\end{align}
We visualize in  \cref{fig:splash-figure} the results of sampling with \cref{eq:retrieval-score}. This definition is motivated by the following proposition, which bounds the distance of \cref{eq:retrieval-score} from the optimal.
\vspace{-7cm}
\begin{proposition}
Assume that $s_{\theta}$ is Lipschitz in $\theta$ and $c$. Let $s_{\theta_0 + \Delta \theta_1}(x_t, t, c)$ be the optimal solution to \cref{eq:loss-1} and let $\Dr$ the private samples retrieved using $c_{\text{test}}$. Then 
\vspace{-3pt}
\begin{align}
\|s_{\theta_1}(x_t, t, c)-\hat{s}_{\theta_0}(x_t, t, c_{\text{test}})\| & \leq \nonumber\\
&\hspace{-6em}  l_{\theta}\|\Delta \theta_1\| + l_{c} \Big\|c_{\text{test}} - \frac{1}{m}\sum_{x_i \in \Dr}\operatorname{CLIP}(x_i)\Big\|
\nonumber
\end{align}
\label{prop:prompt}
\end{proposition}
\vspace{-16pt}
Above result shows that we can approximate the optimal  diffusion model trained on retrieved data using the engineered prompt $\frac{1}{m} \sum_{x_i \in \Dr}\operatorname{CLIP}(x_i)$, which only requires computing the CLIP embeddings of the retrieved images.
Combining \cref{prop:mixture-of-score} and \cref{prop:prompt}, we finally obtain an expression for the score function of retrieval-augmented mixture of distributions, which we call \textit{retrieval-mixture-score}:
\begin{align}
&s_{RAG}(x_t, t, c_{\text{test}}; \Dp) \triangleq \hat{w_0} s_{\theta_0}(x_t, t, c_{\text{test}}) \nonumber \\
&\hspace{4em}+ \hat{w_1} s_{\theta_0}\Big(x_t, t, (1/m)\sum_{x^i \in \Dr}\text{CLIP}(x^i)\Big)
\label{eq:retrieval-score-complete}
\end{align}
\cref{eq:retrieval-score-complete} allows us any pre-trained CLIP-based diffusion model to generate retrieval augmented samples without any  further changes (see \cref{fig:splash-figure}). 
In \cref{fig:tifa-table} we show that \cref{eq:retrieval-score-complete} improves text-to-image alignment (TIFA goes from 81.4 to 82.21). 
Parallelly, the retrieval-mixture score function already has immediate application to privacy, since it makes it trivial to unlearn examples contained in $\Dp$ in constant time: these samples are not used to train any parameter, and hence can be forgotten by simply removing them from disk. However, samples retrieved at inference time can still leak private information, which tackle this next.

\begin{algorithm}[t]
\KwInput{$\nabla_{x_t} \log \int q_t(x_t|x_0) q^{(1)}(x|c) dx_0$, $\nabla_{x_t} \log \int q_t(x_t|x_0) q^{(2)}(x|c) dx_0$, T, N, $c_{\text{test}}$}
\KwOutput{$x_0$}
$x_T \sim \mathcal{N}(0, I)$

\For{$t = T \cdots 0$}{
\For{$i = 1 \cdots N$}{

$x_t = x_t + \dfrac{\epsilon_t}{2} \cdot \dfrac{1}{2} \Big(\nabla_{x_t} \log \int q_t(x_t|x_0) q^{(1)}(x|c_{\text{test}}) dx_0 + \nabla_{x_t} \log \int q_t(x_t|x_0) q^{(2)}(x|c_{\text{test}} dx_0)\Big) + \sqrt{\epsilon_t} z$
where $z \sim \mathcal{N}(0,I)$
}
$x_{t-1} = x_{t}$
}
\caption{CPR-KL}
\label{algo:cpr-kl}
\end{algorithm}
\section{Copy-Protected Generation}
\label{sec:cpr}
In this section, we will provide algorithms for copyright protected generation  --- in the Near-Access Freeness sense of \cite{vyas2023provable} --- using our mixed-privacy RAG method. 

\paragraph{Near-Access Freeness}
Let $\Dp$ be a set of private samples, whose information we want to protect, and let  $\Delta$ be a divergence measure between probability distributions, such as the KL-divergence $\Delta_{KL}$ or thr max-divergence $\Delta_{\text{max}}$ (that is, the Renyi Divergence as $\alpha \rightarrow \infty$). Let $\safe: \Dp \rightarrow \mathcal{M}$ be a function which maps a sample $x^i \in \Dp$ to a generative model trained without using that $x^i$. 
The Near Access-Free (NAF) criteria is defined as:
\begin{definition}[NAF Definition 2.1 in \cite{vyas2023provable}]
We say that a generative model $p(x|c)$ is $k_c$-near access-free (or $k_c$-NAF) on a prompt $c$ with respect to $\Dp$, and $\Delta$, $\safe$, if for all $x^i \in \Dp$ we have
$\Delta \Big(p(x|c) || \safe_{x^i}(x|c) \Big) \leq k_c$.
\label{def1}
\end{definition}
In practice, $\safe_{x^i}$ can be a model trained with \texttt{leave-one-out}, or be \texttt{sharded-safe} \cite{vyas2023provable}, or simply be the safe core diffusion model 
 $s_{\theta_0}(x_t, t, c)$. 
The above definition says that 
to perform safe generation the output sample must be close in distribution to a model which did not have access to the private samples in $\Dp$.
\begin{algorithm}[t]
\KwInput{$c_{\text{test}}$, $\widetilde{s}(x_t, t, c; q^{(1)})$, $\widetilde{s}(x_t, t, c; q^{(2)})$, $J$,  $\text{reverse-update}(x_t, s_t)$}
\KwOutput{$x_0$}
$x_T \sim \mathcal{N}(0, I)$

\For{$t = T \cdots, 0$}{
\If{$t \in J$}
{
$s(x_t,t,c_{\text{test}}) = \widetilde{s}(x_t, t, c_{\text{test}}; q^{(2)})$
}

\ElseIf{$t \not\in J$}
{
$s(x_t,t,c_{\text{test}}) = \widetilde{s}(x_t, t, c_{\text{test}}; q^{(1)})$
}
$x_{t-1} = \text{reverse-update}(x_t, s(x_t,t,c_{\text{test}}))$
}
\caption{CPR-Choose}
\label{algo:cpr-int}
\end{algorithm}

\begin{figure*}[t]
\centering
\includegraphics[width=0.95\textwidth]{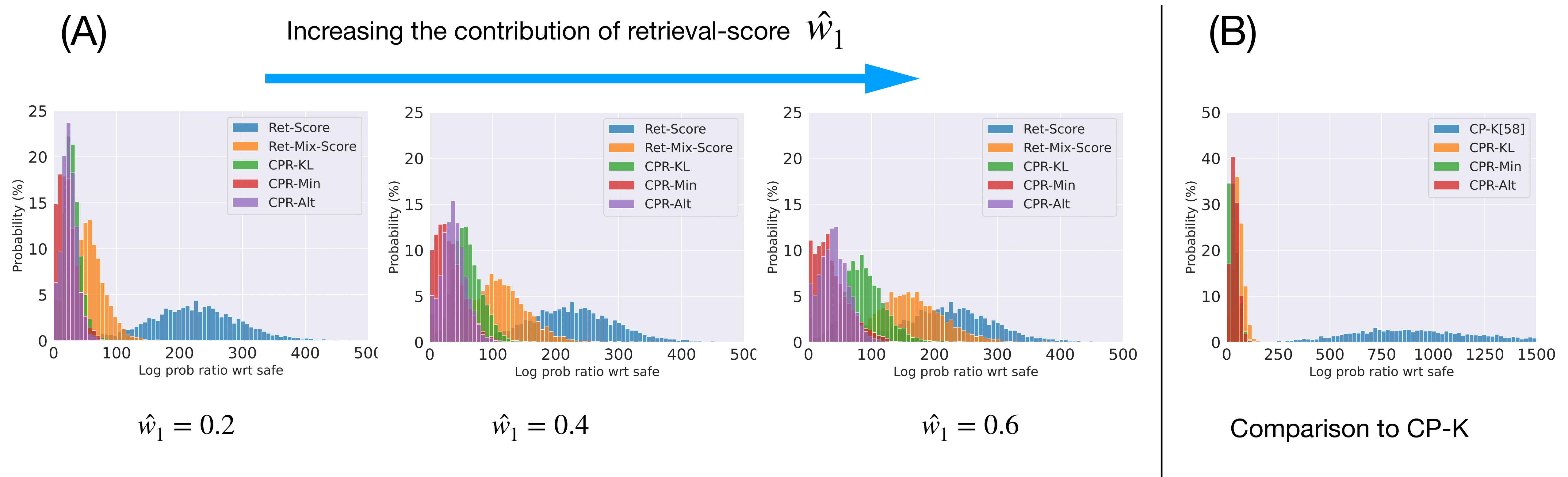}
\vspace{-10pt}
\caption{\textbf{(A)} We plot the histogram of $\Delta_{\text{max}} = \log \dfrac{p(x|c)}{\safe(x|c)}$ as we vary the contribution of the retrieval-score ($\hat{w_1}$ in \cref{eq:retrieval-score-complete}). We use $\hat{w_1}$ as a user tunable parameter which controls the amount of bits the generated images are different from $\safe$. We show that as we reduce $\hat{w_1}$, empirical $k_c$ (max value on the x-axis with non-zero probability) decreases. \textbf{(B)} Comparison to baseline, \cite{vyas2023provable}, with k=1500 using rejection sampling. Smaller k leads to slow generation which is evident from the distribution.}
\label{fig:log-probs}
\end{figure*}

\subsection{CPR-KL}
We first report here Theorem 3.1 from \cite{vyas2023provable} which provides a simple procedure to generate NAF-protected samples with respect to KL-divergence.

\begin{theorem} \label{thm:NAF_KL}
(Theorem 3.1 \cite{vyas2023provable})
Given a dataset $\widetilde{D}$, and copyrighted samples $\mathcal{C} \in \widetilde{D}$, split $\widetilde{D}$ into two disjoint shards $D_1, D_2$, and train two generative models $q^{(1)}, q^{(2)}$ on each respectively. Given the two models return a new model which satisfies $k_c$-NAF wrt $\Delta_{KL}$

\begin{align}
p(x|c) = \dfrac{\sqrt{q^{(1)}(x|c) q^{(2)}(x|c)}}{Z(c)}
\label{eq:prod-sampling}
\end{align}
where $k_c = -2 \log{(1 - \text{H}^2(q^{(1)}(x|c), q^{(2)}(x|c)))}$, H is the Hellinger distance.
\end{theorem}
However, for diffusion models we do not have access to $q^{(1)}$ and $q^{(2)}$, but only to the scores $\nabla_{x_t} \log \int q_t(x_t|x_0) q^{(1)}(x|c) dx_0$ and $\nabla_{x_t} \log \int q_t(x_t|x_0) q^{(2)}(x|c) dx_0$ respectively, where $q_t(x_t|x_0)$ is a variance preserving Gaussian distribution. 

We therefore extend \cref{thm:NAF_KL} to generative models by extending it to models' scores.

Given score functions, we define the CPR-KL algorithm in \cref{algo:cpr-kl} where we average the two scores at every step during backward diffusion using Langevin Dynamics \cite{du2023reduce,song2019generative,welling2011bayesian,chen2014stochastic,neal2001annealed}. 
In the following result we prove that sampling using \cref{algo:cpr-kl} indeed ensures $k_c-$NAF.

\begin{lemma}
Let $x_0$ be the output of \cref{algo:cpr-kl}. 
Under certain regularity conditions (see Supplementary Material), $x_0$ is $k_c-$NAF \wrt $\safe$, $\mathcal{C}$, $\Delta_{\text{KL}}$.
\label{lemma:cpr-kl}
\end{lemma}

By the previous result, \cref{algo:cpr-kl} enables us to generate samples from \cref{eq:prod-sampling}, as $T, N$ increases and $\epsilon_t$ decreases. However, in practice we do not have access to the optimal scores, but instead approximations which use DNNs. 
In our setting we shall consider having the safe model $s_{\theta_0}(x_t,t,c)$, and the RAG score on the private datastore $s_{RAG}(x_t, t, c_{\text{test}}; \Dp)$.
In practice, although combining the two scores as in \cref{algo:cpr-kl} can produce better results, it also doubles the computation cost at inference time. 
To circumvent this we now describe CPR-Choose (\cref{algo:cpr-int}) which approximates \cref{algo:cpr-kl} without increasing computational complexity.

\begin{figure*}[t]
\centering
\includegraphics[width=0.9\textwidth]{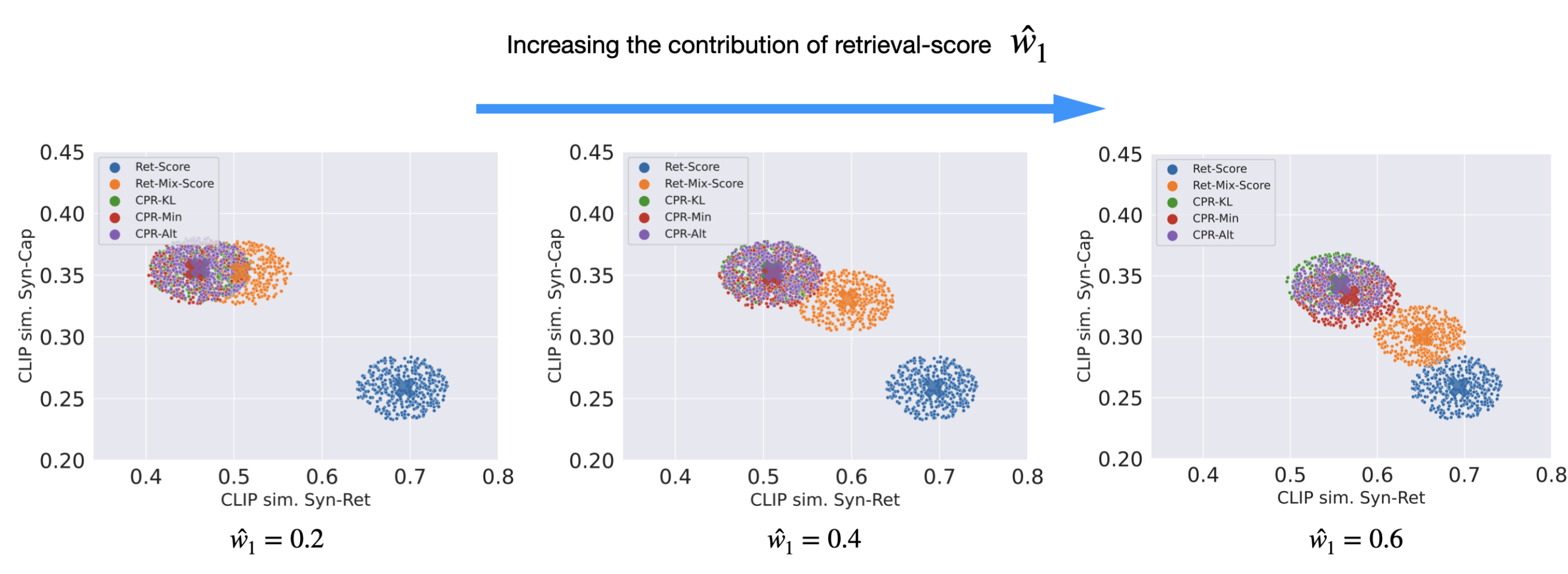}
\caption{Concept similarity with CPR: In this figure we show the CLIP similarity between CPR generated images and the textual prompt (Syn-Cap) and the retrieved images (Syn-Ret) respectively. We show that while the CPR generated image preserves the concept presented in the textual prompt (their similarity with the caption is high), they do not copy the private retrieved images (their similarity with the retrieved samples is low).}
\label{fig:similarity}
\end{figure*}

\subsection{CPR-Choose}
We now propose another CPR algorithm which does not incur in higher computational cost of CPR-KL. First we recall a result on the likelihood estimation of samples with MMSE denoisers, then we show how to use it to define an efficient NAF algorithm \wrt $\Delta_{\text{max}}$.  

\textbf{Estimating sample likelihood with MMSE denoiser}
Recently, \cite{kong2023information, kong2023interpretable} provided a simple method for estimating the probability of individual samples by computing the Minimum mean square error (MMSE) using pre-trained text-to-image diffusion models. Let $x_t = \gamma_t x_0 + \sigma_t \epsilon$, $x_0 \sim p(x_0|c)$ where $\epsilon \sim \mathcal{N}(0, I)$, and $\alpha(t) = \log \dfrac{\gamma^2_t}{\sigma^2_t}$ be the log SNR. Then the MMSE denoiser for a distribution $p$ can defined as: 
\begin{align}
&\widetilde{s}(x_t, t, c; p) \triangleq \text{argmin}_{s(\cdot)}\mathbb{E}_{p(x_0|c), \epsilon}\|\epsilon - s(x_t,t,c)\|^2 \nonumber \\
&= \mathbb{E}_{p(x_0|x_t, c)} \Big[ \dfrac{x_t - \gamma_t x_0}{\sigma_t} \Big]
\label{eq:mmse-optimal}
\end{align}
Using the MMSE denoiser, \cite{kong2023information,kong2023interpretable} provide a simple expression for estimating the log probability of $x_0$.
\begin{align}
\log p(x_0|c) = -\int \mathbb{E}_{\epsilon}\|\epsilon - \widetilde{s}(x_t, t, c; p)\|^2 \alpha'(t) dt + const
\label{eq:prob-mmse}
\end{align}
where $\alpha'(t)$ is the time-derivative of $\alpha(t)$. Note that $\widetilde{s}(x_t, t, c; p)$ is also equivalent to the diffusion score we obtained in the previous sections. 

This result shows that to obtain NAF \wrt $\Delta_{\text{max}}=\log p(x|c)/\safe(x|c)$, all we need to do is bound the difference in MMSE at each time step $t$. We can bound this by choosing $p(x|c)=\safe(x|c)$ for majority of $t$, while using $\Dp$ intermittently for remaining $t$.

Using these results we will provide another algorithm for copy-protected generation. Let $q^{(1)}, q^{(2)}$ be the models obtained using $D_1, D_2$ respectively. And assume that we shard the total data in such a way that $D_1$ contains the safe data, while $D_2$ contains the copy-protected data. \textit{We let $q^{(1)}$ be our $\safe$-model}. In practice, we will have access to the score function or the MMSE denoiser, $\widetilde{s}(x_t, t, c; q^{(1)})$, $\widetilde{s}(x_t, t, c; q^{(2)})$.

\textbf{NAF $\Delta_{max}$ algorithm}
Let $J = \{[t_i, t_{i+1}] | t_{i+1} \leq t_{i+2},  i \in \{0,2,4,\cdots,N\}, t_0 \geq 0,t_{N+1} < \infty, N < \infty \}$ be a subset of disjoint time-intervals on the real line. Using the set $J$, let us define a new distribution,
\begin{align}
\widetilde{q}(x_0|c,t) = q^{(1)}(x_0|c) \mathbbm{1}_{t \not\in J} + q^{(2)}(x_0|c) \mathbbm{1}_{t \in J}
\label{eq:time-dep-dist}
\end{align}
This new distribution is a time-dependent, which essentially selects a distribution at time $t$ to sample $x_t$. The benefit of such an approach is that it enables the user to select one of the two model during backward diffusion at each time-step, which is similar to \cite{balaji2022ediffi} which has shown to empirically improve generation quality. Towards this end we have the following result,

\begin{proposition}
Let $\widetilde{s}(x_t, t, c; \widetilde{q})$ be the MMSE denoiser for \cref{eq:time-dep-dist}, then we can show that 
\begin{align}
\widetilde{s}(x_t, t, c; \widetilde{q}) = \widetilde{s}(x_t, t, c; q^{(1)}) \mathbbm{1}_{t \not\in J} + \widetilde{s}(x_t, t, c; q^{(2)}) \mathbbm{1}_{t \in J} \nonumber
\end{align}
\label{prop:time-dep-mmse-den}
\end{proposition}
\vspace{-25pt}
This result states the fact that optimal MMSE denoiser for \cref{eq:time-dep-dist} will choose one of the two denoisers depending on the time-step, where the choice of $J$ can be completely user dependent. Using these observations we propose interval based CPR algorithm (CPR-Choose), \cref{algo:cpr-int}

\begin{lemma}
Let $x_0$ be the output of \cref{algo:cpr-int}. 
Under certain regularity conditions (see Supplementary Material), $x_0$ is $k_c-$NAF \wrt $\safe$, $\mathcal{C}$, $\Delta_{\text{max}}$.
\label{lemma:choose}
\end{lemma}

\begin{figure*}[t]
\vspace{-10pt}
\centering
\includegraphics[width=0.8\textwidth]{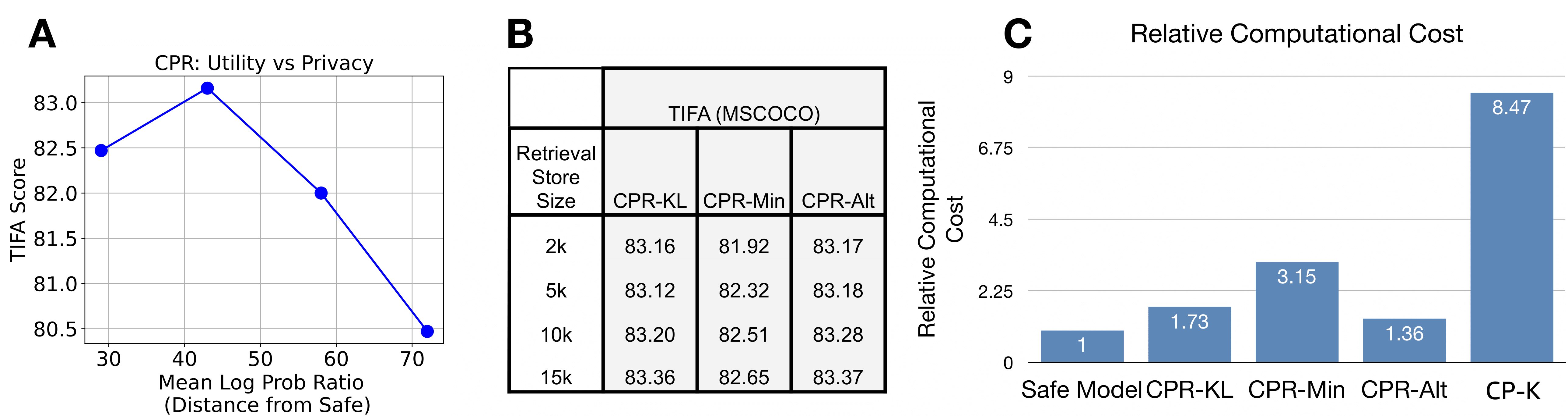}
\caption{(a) Plot of the utility (generation quality) for increasing values of copyright protection, on samples from the MS-COCO dataset. (b) The TIFA score of CPR increases as the size of the retrieval dataset grows. (c) Computational costs of CPR (ours) and CP-K\cite{vyas2023provable} compared to the base model.}
\label{fig:rebuttal_exp}
\end{figure*}

\vspace{-10pt}
\subsubsection{Time-Discretization} 
\label{sec:time-discrete}
Often in practice we model the diffusion process using a discrete markov chain \cite{ho2020denoising,song2020denoising} whose continuous limit is SDE\cite{song2020score}. For discrete markov chains discrete in $t$ we can denote the output of models $q^{(1)}$ ($\safe$-model), $q^{(2)}$ using the entire trajectory, $\{x_0, \cdots, x_T\}$) \cite{vyas2023provable}. The set of intervals $J$ becomes a set of discrete time-steps. During backward diffusion, at each $t$ the user can use one of the two models to generate the score for updating $x_t$.  Depending on the choice of $J$, we can generate completely safe images ($J$ to be empty) or no protection ($J$ is the entire domain of $t$).This leads to two CPR-Choose algorithms, depending on the choice of J.

\paragraph{CPR-Min}\label{cpr-min}
In this setting, we choose $J$ such that at each $t$, we choose the model with the larger MSE, which can be considered as choosing the worst model at each $t$. This will generate samples from a distribution which approximates the \textit{minimum} of the two distributions. Under certain conditions we can show that this algorithm is NAF protected (In the appendix). This in intuitive because, for time-stamps when we choose $q^{(1)}$ (which is the $\safe$ model), we incur no loss for $\Delta_{\text{max}}$, and it is only for the remaining terms that we need to bound $\Delta_{\text{max}}$.

\paragraph{CPR-Alt}\label{cpr-alt}
Similarly, we can choose $J$ to \textit{alternate} between the two models by choosing $q^{(2)}$ (private model)  at regular intervals, like \eg every $\widetilde{t}$ steps, or in the most simplest case, in an alternating fashion. Using this approach, we will only need to compute the $\Delta_{\text{max}}$ at every $\widetilde{t}$ steps to bound $k_c$.

In our experiments, we will let $q^{(1)}$ be the $s_{\theta_0}(x_t,t,c)$ which is trained on the safe-core data, while $q^{(2)}$ be the $s_{\text{RAG}}(x_t,t,c; \Dp)$ which uses the private data at inference using retrieval.

\section{Experiments}
\label{sec:exps}
We use the Stable-Diffusion 2.1 model \cite{rombach2022high} as our safe base model, and use the Stable-Diffusion unCLIP model \cite{rombach2022high,ramesh2022hierarchical} (without the prior model) as our retrieval-score model. Using the unCLIP model enables better control of the generation with the retrieved images $\Dr \subset \Dp$. We use ~top 2k samples (based on the aesthetic score) from MSCOCO \cite{lin2014microsoft} as our private data store and use the TIFA score \cite{hu2023tifa} to measure the text-image alignment and quality.

\paragraph{Improved text-to-image alignment}
Retrieval is often used to improve the text-to-image alignment of the diffusion model. In \cref{fig:tifa-table}, we use TIFA benchmark to evaluate the alignment of different methods. We observe that retrieving images from the data store indeed improves the alignment from 81.4 to 83.17. Interestingly, CPR regularizes the inference, resulting in even better TIFA (with protection).

\paragraph{Comparing privacy leakage}
In \cref{fig:log-probs}, we plot the $\Delta_{\text{max}}$ (whose upper bound is $k_c$) for various methods against $\safe$ (on images generated with TIFA prompts). We use the control parameter $\hat{w}_1$ (\cref{eq:retrieval-score-complete} to vary the retrieval contribution. We show that increasing $\hat{w}_1$, makes the model generate more similar images to $\Dp$, resulting in larger $\Delta_{\text{max}}$ (log prob. ratio \wrt safe). 
This is unlike the CP-$\Delta$ \cite{vyas2023provable} which does not allow the user to tune the NAF constant $k_c$. 
We also compare with CP-K \cite{vyas2023provable}, which uses rejection sampling on the outputs generated by a Stable Diffusion model fine-tuned on the private database $\Dp$.
We set k=1500, and observe that $\log p(x|c)/\safe(x|c)$ is almost uniformly distributed, which results in much slower (5-10x) rejection sampling for the same privacy level as our CPR algorithms.

\paragraph{Concept similarity with CPR}
In \cref{fig:similarity}, we plot the CLIP-score between the image generated using TIFA prompts (Syn in \cref{fig:similarity}) and the input captions (Cap in \cref{fig:similarity}), retrieved images from $\Dp$ (Ret in \cref{fig:similarity}) respectively. We show that CPR reduces the similarity between the synthesized images and the retrieved images, while improving the similarity to the textual prompts. This implies that CPR generates images corresponding to the concept present in the prompt (with the help of the retrieved image), but ensures that the synthesized image is different from the retrieved image (prevents copying/memorization). 

\paragraph{Ablations}
In \cref{fig:rebuttal_exp} we provide additional experiments where we ablate the size of the retrieval store, show the privacy utility trade-off, and compare the computations cost of various methods.

\section{Discussion and Limitations}
\label{sec:disc}

\paragraph{Relation between $k_c$ and retrieval function}
The NAF bound $k_c$ relates to the private data store through the retrieval function, which in our case is the $L_2$ distance between the CLIP embeddings. 
Functions that retrieve images which explain the concept underlying the $c_{\text{test}}$ instead of its exact expression, can further help in improving privacy.

\paragraph{Classifier-free guidance for privacy protected generation}
We can redefine the expression in \cref{eq:prod-sampling}, to represent a more general form like $p(x|c) \propto q_1^{\alpha}(x|c)q_2^{1-\alpha}(x|c)$, which when substituted with appropriate $\alpha$ provides the expression for the classic classifier free guidance (CFG) \cite{ho2022classifier}, which implies that replacing the marginal in CFG with a safe model, and using the RAG model in place of the conditional results in private generation with appropriate scaling of $k_c$. Thus CFG with appropriate model selection can be considered a good candidate for NAF generation.

\paragraph{Unlearning, adapters, and RAG} A direct consequence of copied generation is the request to remove the appropriate training samples from the dataset (in our case $\Ds$). Such unlearning requests can be efficiently handled by our CPR framework as it allows for cost free removal of private samples. However, in certain settings, if the private data store contains out-of-distribution (OOD) examples, simply using $\cref{eq:retrieval-score-complete}$ may not be enough to obtain high fidelity images. In such situations we may train separate adapters \cite{jia2022visual,sohn2023visual,dukler2023introspective,dukler2023safe,hu2021lora} corresponding to the OOD samples (a subset of $\Ds$. Hence at inference, we would first retrieve a private adapter, and then a set of samples from $\Ds$. Upon a forgetting request, we discard both the adapter, and the samples in $\Ds$.

\paragraph{Limitations}
One of the major limitation of diffusion model based methods is the inability to compute the exact probability values (this is not the case for auto-regressive or flow based models). For instance, in \cref{prop:mixture-of-score}, $\hat{w_0}$, or even the computation of the true NAF parameter depends on the actual probability values.

{\small
\bibliographystyle{ieeenat_fullname}
\bibliography{main}}

\clearpage
\newpage
\onecolumn

\appendix
\begin{center}
\begin{Large}
CPR: Retrieval Augmented Generation for Copyright Protection
\end{Large}
\end{center}
\begin{center}
\begin{large}
Supplementary Material
\end{large}
\end{center}

\section{Proofs of the Propositions and Lemmas}

\subsection{\cref{prop:mixture-of-score}}
\begin{proof} of \cref{prop:mixture-of-score}.
\begin{align}
\nabla_{x_t} \log p_t(x_t|c) &=  \nabla_{x_t} \log \int p_t(x_t|x_0)\big[w_0p_D(x_0|c) + w_1p_{D_\text{retr}}(x_0|c)\big]dx_0 \nonumber \\
&= \dfrac{1}{\int p_t(x_t|x_0)\big[w_0p_D(x_0|c) + w_1p_{D_\text{retr}}(x_0|c)\big]dx_0} \Big[\nabla_{x_t} \int p_t(x_t|x_0) w_0p_D(x_0|c) dx_0 \nonumber \\ &+  \nabla_{x_t} \int p_t(x_t|x_0) w_1p_{D_\text{retr}}(x_0|c) dx_0\Big] \nonumber \\
&= \dfrac{1}{p_t(x_t|c)} \Big[\nabla_{x_t} \int p_t(x_t|x_0) w_0p_D(x_0|c) dx_0 +  \nabla_{x_t} \int p_t(x_t|x_0) w_1p_{D_\text{retr}}(x_0|c) dx_0\Big] \nonumber \\
&= \dfrac{1}{p_t(x_t|c)} \Big[w_0\int p_t(x_t|x_0) p_D(x_0|c) dx_0 \nabla_{x_t} \log \int p_t(x_t|x_0) p_D(x_0|c) dx_0 \nonumber \\ &+  w_1\int p_t(x_t|x_0) p_{D_\text{retr}}(x_0|c) dx_0 \nabla_{x_t} \log \int p_t(x_t|x_0) p_{D_\text{retr}}(x_0|c) dx_0\Big] \nonumber \\
&= \dfrac{w_0\int p_t(x_t|x_0) p_D(x_0|c) dx_0}{p_t(x_t|c)} \nabla_{x_t} \log \int p_t(x_t|x_0) p_D(x_0|c) dx_0 \nonumber \\ &+  \dfrac{w_1\int p_t(x_t|x_0) p_{D_\text{retr}}(x_0|c) dx_0}{p_t(x_t|c)}  \nabla_{x_t} \log \int p_t(x_t|x_0) p_{D_\text{retr}}(x_0|c) dx_0\ \nonumber
\end{align}
\end{proof}

\subsection{\cref{prop:prompt}}
\begin{proof} of \cref{prop:prompt}.
Let $s_{\theta_1}(x_t, t, c) \triangleq s_{\theta_0 + \Delta \theta_1}(x_t, t, c)$ be the optimal solution to the retrieval optimization problem. We use $\operatorname{CLIP}$ embeddings of the retrieved images for generation, and bound its difference from the optimal.
\begin{align}
\|s_{\theta_1}(x_t, t, c)-\hat{s}_{\theta_0}(x_t, t, c_{\text{test}})\| &=
\|s_{\theta_1}(x_t, t, c)-s_{\theta_0}\Big(x_t, t, \frac{1}{m} \sum_{x_i \in \Dr}\text{CLIP}(x_i)\Big)\| \nonumber \\
&= \|s_{\theta_1}(x_t, t, c) - s_{\theta_0}(x_t, t, c) + s_{\theta_0}(x_t, t, c) -  s_{\theta_0}\Big(x_t, t, \frac{1}{m} \sum_{x_i \in \Dr}\text{CLIP}(x_i)\Big)\| \nonumber \\
&\leq \|s_{\theta_1}(x_t, t, c) - s_{\theta_0}(x_t, t, c)\| + \|s_{\theta_0}(x_t, t, c) -  s_{\theta_0}\Big(x_t, t, \frac{1}{m} \sum_{x_i \in \Dr}\text{CLIP}(x_i)\Big)\| \nonumber \\
&\leq \|s_{\theta_0 + \Delta \theta_1}(x_t, t, c) - s_{\theta_0}(x_t, t, c)\| + \|s_{\theta_0}(x_t, t, c) -  s_{\theta_0}\Big(x_t, t, \frac{1}{m} \sum_{x_i \in \Dr}\text{CLIP}(x_i)\Big)\| \nonumber \\
&\leq l_{\theta} \|\Delta \theta_1\| + l_{c} \|\frac{1}{m} \sum_{x_i \in \Dr}\text{CLIP}(x_i)\|
\end{align}
\end{proof}

\subsection{\cref{lemma:cpr-kl}}
\begin{proof} of \cref{lemma:cpr-kl}. \cite{vyas2023provable} proved in Theorem 3.1, that sampling from \cref{eq:prod-sampling} produces samples which are copy-protected. In \cref{algo:cpr-kl}, we sample using the score function: $ 0.5 (\nabla_{x_t} \log \int q_t(x_t|x_0) q^{(1)}(x|c) dx_0 + \nabla_{x_t} \log \int q_t(x_t|x_0) q^{(2)}(x|c) dx_0$, which smoothly interpolates between $\mathcal{N}(0,I)$ at $t=T$, and \cref{eq:prod-sampling} at $t=0$. We need to show that using Langevin based backward diffusion in \cref{algo:cpr-kl} indeed generates samples from the desired distribution. The convergence results for Langevin dynamics have been well studied in practice \cite{neal2001annealed,cheng2018convergence,erdogdu2021convergence,vempala2019rapid}, \cite{roberts1996exponential} has shown that Langevin dynamics converge exponentially fast to the distribution estimated by the gradients. Theorem 2.1 from \cite{roberts1996exponential} provides the result on the convergence of Langevin dynamics in continuous time. For the sake of completeness we will extend the results from \cite{yang2022convergence} to show that \cref{algo:cpr-kl} generates samples from \cref{eq:prod-sampling}. 

We will re-state the assumptions from \cite{yang2022convergence}, for a distribution $\nu_t(x_t)$, and score estimator $s_t(x_t)$. In our case $\nu_t(x_t) = 0.5 (\nabla_{x_t} \log \int q_t(x_t|x_0) q^{(1)}(x|c) dx_0 + \nabla_{x_t} \log \int q_t(x_t|x_0) q^{(2)}(x|c) dx_0)$, and $s_t(x_t)$ is the average of the safe diffusion flow and retrieval mixture score.
\begin{enumerate}
\item LSI: For any probability distribution $\rho$, $C_0>0$, $\int \rho_t \log \dfrac{\rho_t}{\nu_t}dx \leq \dfrac{1}{2C_0}\int \rho_t \Big \|\nabla \log \dfrac{\rho_t}{\nu_t} \Big\|dx$
\item L-Smoothness: $-\log \nu_t$ is L-smooth
\item Lipschitz score estimator: $s_t(x_t)$ is $L_s$-lipschitz
\item MGF error assumption: $M_t = \sqrt{\mathbb{E}_{\nu_t}[\exp{r\|\nabla \log \nu_t(x_t) - s_t(x_t)\|^2}]} \leq \infty$
\end{enumerate}
Then from Theorem 1 in \cite{yang2022convergence} we know that
\begin{align}
\operatorname{KL}(\rho_t(x_t)||\nu_t(x_t)) \leq \exp{(-\dfrac{1}{4}C_0hN)} \operatorname{KL}(\rho_{t+1}(x_{t+1})||\nu_{t+1}(x_{t+1})) + C_1 \epsilon_t + C_2 M_t
\label{eq:kl-recur}
\end{align}
where $N$ is from the \cref{algo:cpr-kl}, $C_1 = O(\dfrac{dLL^2_s}{C_0})$, $C_2 = \dfrac{16}{3}$. \cref{eq:kl-recur} result is the obtain by running the inner loop in \cref{algo:cpr-kl}. Using the previous equation recursively for \cref{algo:cpr-kl}, we obtain that, 
\begin{align}
\operatorname{KL}(\rho_0(x_0)||\nu_0(x_0)) &\leq \exp{(-\dfrac{1}{4}C_0hNT)} \operatorname{KL}(\rho_{T}(x_{T})||\nu_{T}(x_{T})) \nonumber \\
&+ \sum_{t=1}^T \exp{(-\dfrac{1}{4}C_0hN(T-t))} \epsilon_t C_1  + \sum_{t=1}^T \exp{(-\dfrac{1}{4}C_0hN(T-t))} M_t C_1
\label{eq:kl-final}
\end{align}
where $\nu_0(x_0)$ is the distribution in \cref{eq:prod-sampling}. Since we use DNNs with sufficient capacity, we can assume that $M_t \rightarrow 0$, then as $\epsilon_t \rightarrow 0$, and $T \rightarrow \infty$, we have that $\operatorname{KL}(\rho_0(x_0)||\nu_0(x_0)) \rightarrow 0$, which implies that \cref{algo:cpr-kl} generates samples from \cref{eq:prod-sampling}.
\end{proof}

\subsection{\cref{prop:time-dep-mmse-den}}
\begin{proof} of \cref{prop:time-dep-mmse-den}
Let $\widetilde{s}(x_t, t, c; \widetilde{q}) = \mathbb{E}_{\widetilde{q}(x_0|x_t, c)} \Big[ \dfrac{x_t - \gamma_t x_0}{\sigma_t} \Big]$, where $\widetilde{q}(x_0|c,t) = q^{(1)}(x_0|c) \mathbbm{1}_{t \not\in J} + q^{(2)}(x_0|c) \mathbbm{1}_{t \in J}$.
\begin{align}
\widetilde{s}(x_t, t, c; \widetilde{q}) &= \mathbb{E}_{\widetilde{q}(x_0|x_t, c)} \Big[ \dfrac{x_t - \gamma_t x_0}{\sigma_t} \Big] \nonumber \\
&= \int \widetilde{q}(x_0|x_t, c) \Big[ \dfrac{x_t - \gamma_t x_0}{\sigma_t} \Big] dx_0 \nonumber \\
&= \int \Big(q^{(1)}(x_0|c) \mathbbm{1}_{t \not\in J} + q^{(2)}(x_0|c) \mathbbm{1}_{t \in J} \Big) \Big[ \dfrac{x_t - \gamma_t x_0}{\sigma_t} \Big] dx_0 \nonumber \\
&= \int q^{(1)}(x_0|c) \mathbbm{1}_{t \not\in J}  \Big[ \dfrac{x_t - \gamma_t x_0}{\sigma_t} \Big] dx_0 + \int q^{(2)}(x_0|c) \mathbbm{1}_{t \in J} \Big[ \dfrac{x_t - \gamma_t x_0}{\sigma_t} \Big] dx_0 \nonumber \\
&= \widetilde{s}(x_t, t, c; q^{(1)}) \mathbbm{1}_{t \not\in J} + \widetilde{s}(x_t, t, c; q^{(2)}) \mathbbm{1}_{t \in J} \nonumber
\end{align}
\end{proof}

\subsection{\cref{lemma:choose}}
\begin{proof} of \cref{lemma:choose}
We use \cref{prop:time-dep-mmse-den} in \cref{algo:cpr-int} for CPR-generation. Let $q^{(1)}$ be the safe model in accordance with the assumptions in \cref{sec:cpr}. To show that \cref{algo:cpr-int} is NAF, we need to bound $\Delta_{\text{max}}$. To show that $\widetilde{q}(x_0|c,t)$ satisfies NAF we need to bound:
\begin{align}
\log \dfrac{\widetilde{q}(x_0|c)}{q^{(1)}(x_0|c)} &= \int \mathbb{E}_{\epsilon} \|\epsilon - \tilde{s}(x_t, t, c; q^{(1)})\|^2\alpha'(t)dt - \mathbb{E}_{\epsilon} \|\epsilon - \tilde{s}(x_t, t, c; \widetilde{q})\|^2\alpha'(t)dt \nonumber \\
&= \int \mathbb{E}_{\epsilon} \big(\|\tilde{s}(x_t, t, c; q^{(1)})\|^2 - \|\tilde{s}(x_t, t, c; \widetilde{q})\|^2 \big) \alpha'(t) dt \nonumber \\
&= \sum_{j \in J} \int_{t \in j} \mathbb{E}_{\epsilon} \big(\|\tilde{s}(x_t, t, c; q^{(1)})\|^2 - \|\tilde{s}(x_t, t, c; \widetilde{q})\|^2 \big) \alpha'(t) dt \nonumber \\
&= \sum_{j=[t_i, t_{i+1}] \in J} \int_{t \in j} \mathbb{E}_{\epsilon} \big(\|\tilde{s}(x_t, t, c; q^{(1)})\|^2 - \|\tilde{s}(x_t, t, c; q^{(2)})\|^2 \big) \alpha'(t) dt \nonumber \\
&= \sum_{j=[t_i, t_{i+1}] \in J, t' \in j} \mathbb{E}_{\epsilon} \big(\|\tilde{s}(x_t', t', c; q^{(1)})\|^2 - \|\tilde{s}(x_t', t', c; q^{(2)})\|^2 \big) \alpha'(t') (t_{i+1}-t_i) \nonumber \\
&= \sum_{j=[t_i, t_{i+1}] \in J, t' \in j} \mathbb{E}_{\epsilon} \big(\|\tilde{s}(x_t', t', c; q^{(1)})\|^2 - \|\tilde{s}(x_t', t', c; q^{(2)})\|^2 \big) \alpha'(t') (t_{i+1}-t_i) \nonumber \\
&\leq \text{max}_{t' \in J} \mathbb{E}_{\epsilon} \big(\|\tilde{s}(x_t', t', c; q^{(1)})\|^2 - \|\tilde{s}(x_t', t', c; q^{(2)})\|^2 \big) \alpha'(t') \sum_{j=[t_i, t_{i+1}] \in J, t' \in j}  (t_{i+1}-t_i)  \nonumber \\
&= k_c
\end{align}
$J$ is our control parameter in CPR-Choose which controls $k_c$. If a conservative approach is to be followed, then $J$ should be chosen such that $\sum_{j=[t_i, t_{i+1}] \in J, t' \in j}  (t_{i+1}-t_i)$ is small, which bounds $k_c$, the copy-protection leakage. 

\paragraph{CPR-Min, CPR-Alt} In practice we discretize the time-steps of the backward diffusion process. In this setting we protect the entire sequence $\{x_T, \cdots, x_0\}$ instead of protecting only the final prediction $x_0$. The probability of the sequence $\{x_T, \cdots, x_0\}$ is denoted by $\widetilde{q}(x_0|x_1, c) \cdots \widetilde{q}(x_{T-1}|x_T, c) \widetilde{q}(x_T|c)$ using the chain rule of probability. To show that the method satisfies  NAF, we need to bound:
\begin{align}
\log \dfrac{\widetilde{q}(\{x_0, \cdots, x_T\}|c)}{q^{(1)}(\{x_0, \cdots, x_T\}|c)} &= \log \prod_t \dfrac{\widetilde{q}(x_t|x_{t+1},c)}{q^{(1)}(x_t|x_{t+1},c)} \nonumber \\
&= \log \prod_{t \in J} \dfrac{q^{(2)}(x_t|x_{t+1},c)}{q^{(1)}(x_t|x_{t+1},c)} \nonumber \\
&= \sum_{t \in J} \log \dfrac{q^{(2)}(x_t|x_{t+1},c)}{q^{(1)}(x_t|x_{t+1},c)} \nonumber \\
&= \sum_{t \in J} \log \dfrac{\mathcal{N}(x_{t}; \alpha_{1,t} x_{t+1} + \alpha_{2,t} \widetilde{s}(x_{t+1}, t+1, c, q^{(2)}), \sigma^2_tI)}{\mathcal{N}(x_{t}; \alpha_{1,t} x_{t+1} + \alpha_{2,t} \widetilde{s}(x_{t+1}, t+1, c, q^{(1)}), \sigma^2_tI)} \nonumber \\
&= \sum_{t \in J} \dfrac{1}{\sigma^2_t} \Big(\|x_t - \alpha_{1,t} x_{t+1} + \alpha_{2,t} \widetilde{s}(x_{t+1}, t+1, c, q^{(1)})\|^2 \nonumber \\
&- \|x_t - \alpha_{1,t} x_{t+1} + \alpha_{2,t} \widetilde{s}(x_{t+1}, t+1, c, q^{(2)})\|^2 \Big) \nonumber \\
&\leq \text{max}_t \Big(\|x_t - \alpha_{1,t} x_{t+1} + \alpha_{2,t} \widetilde{s}(x_{t+1}, t+1, c, q^{(1)})\|^2 \nonumber \\
&- \|x_t - \alpha_{1,t} x_{t+1} + \alpha_{2,t} \widetilde{s}(x_{t+1}, t+1, c, q^{(2)})\|^2 \Big) \sum_{t \in J} \dfrac{1}{\sigma^2_t} \nonumber \\
&\leq b \sum_{t \in J} \dfrac{1}{\sigma^2_t} \nonumber \\
&= k_c
\end{align}
where $\alpha_{1,t}, \alpha_{2,t}, \sigma^2_t$ are the coefficients using the backward diffusion depending on the choice of sampler, for eg. DDPM \cite{ho2020denoising}, DDIM \cite{song2020denoising}, Langevin dynamics \cite{du2023reduce}, $b$ is an upper bound on the maximum difference between the MSE for the two diffusion processes. Similar to the previous derivation, $\sum_{t \in J} \dfrac{1}{\sigma^2_t}$ through $J$ provides a control knob to the user to control the $\Delta_{\text{max}}$ for copy-protected generation.
\end{proof}

\section{Implementation Details}
We use the Stable diffusion \cite{rombach2022high} and Stable diffusion unCLIP \cite{ramesh2022hierarchical} model for all the experiments in the paper. We use the Stable diffusion model to generate safe flow corresponding to the safe distribution $q^{(1)}$, and the Stable diffusion unCLIP model to generate the retrieval mixture score $q^{(2)}$. We use classifier free guidance with a guidance scale of 7.5 in all the results. We use ~2k samples from the MSCOCO dataset \cite{lin2014microsoft} as our private retrieval data store.

\section{Additional Figures}

\begin{figure*}[h]
\centering
\includegraphics[width=\textwidth]{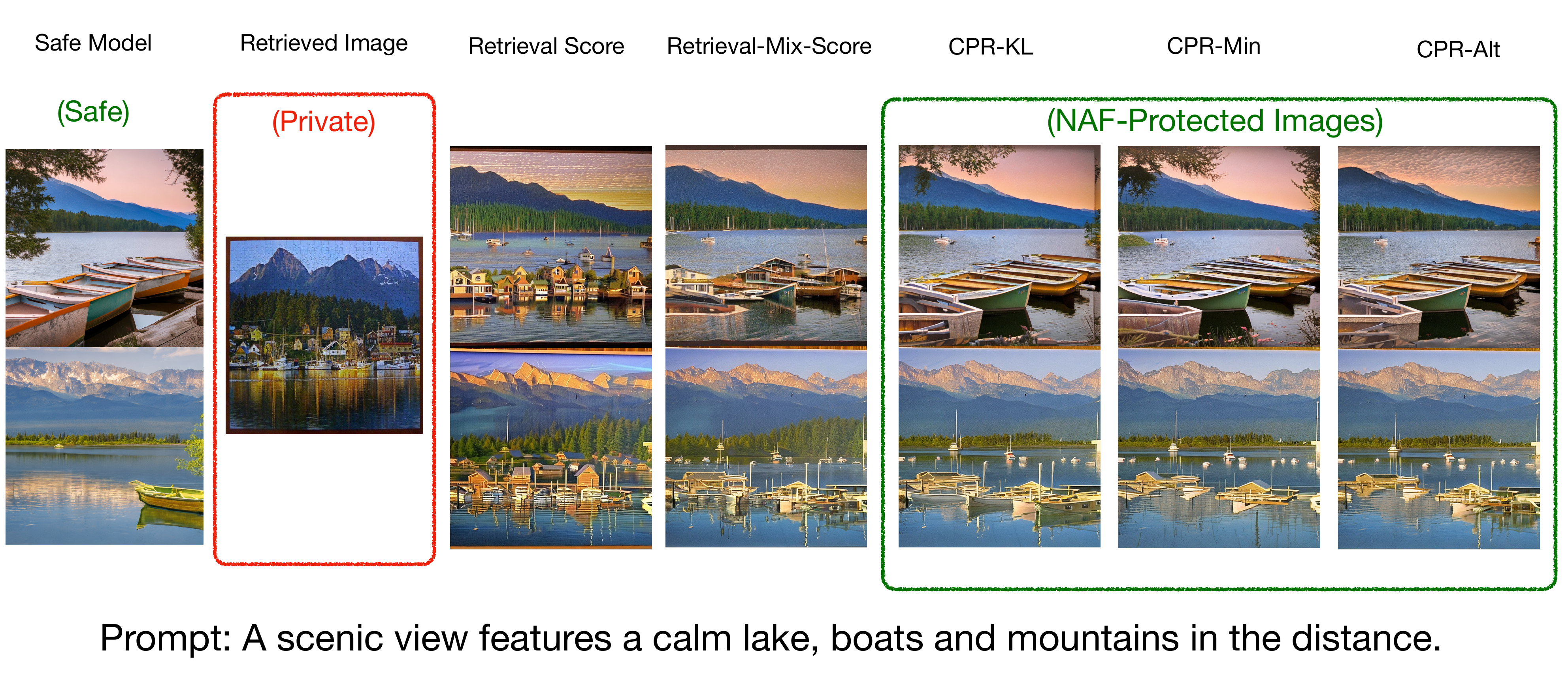}
\caption{}
\end{figure*}

\begin{figure*}[h]
\centering
\includegraphics[width=\textwidth]{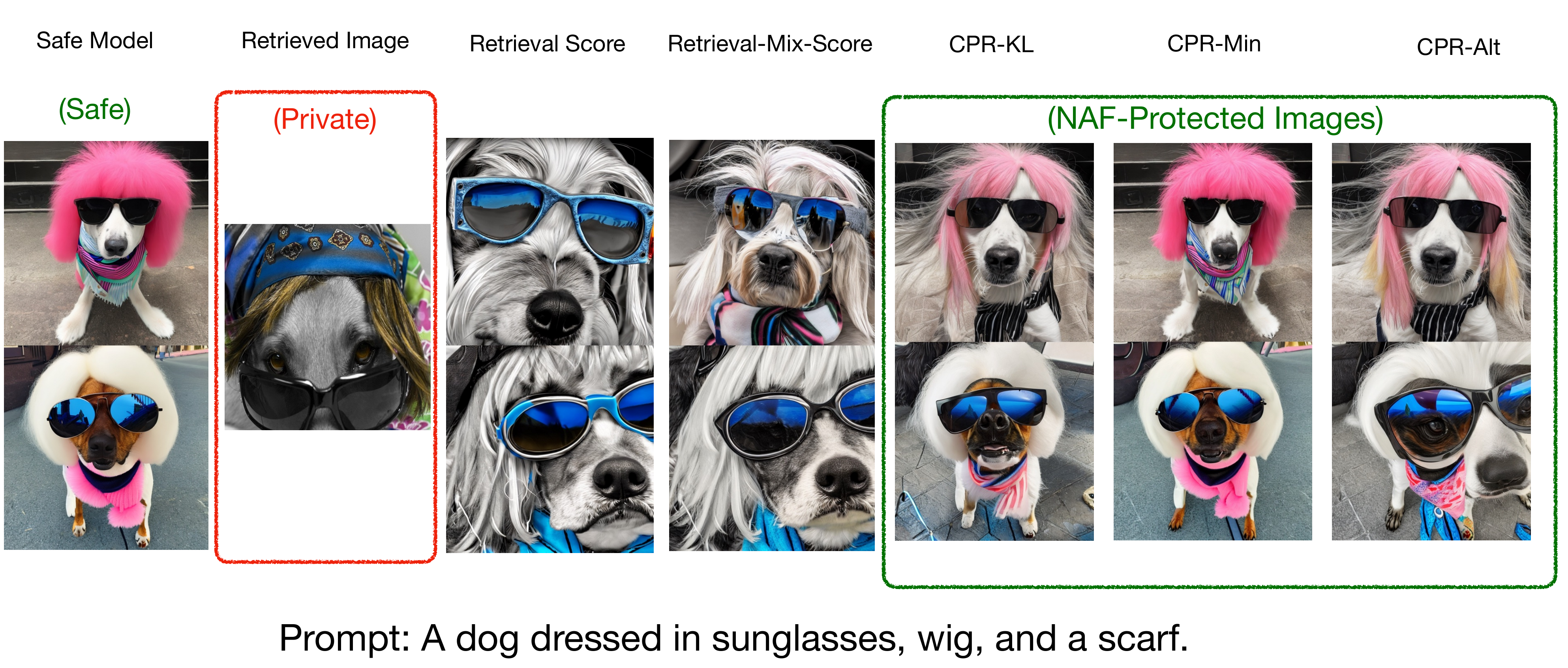}
\caption{}
\end{figure*}

\begin{figure*}[h]
\centering
\includegraphics[width=\textwidth]{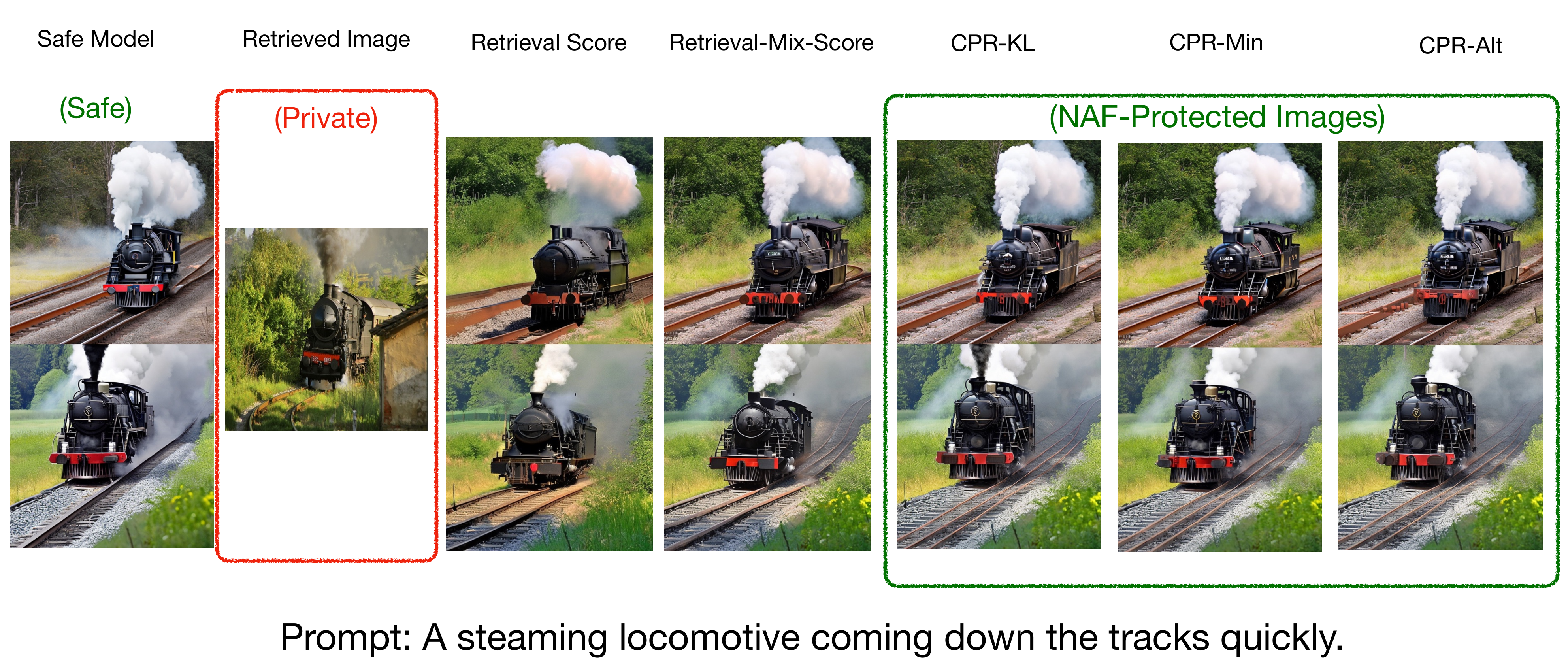}
\caption{}
\end{figure*}

\end{document}